\documentclass[aps,pra,twocolumn,superscriptaddress]{revtex4-1}
\usepackage{amsmath}
\usepackage{cases}
\usepackage{graphicx}
\usepackage{mathtools} % celtic 
\usepackage[T1]{fontenc} % celtic
\usepackage[utf8]{inputenc} % celtic
\usepackage{amsfonts} %celtic
\usepackage{bm}
\usepackage{dsfont}

\usepackage{enumitem}
\usepackage{multirow}
\usepackage{tabularx}
\usepackage{longtable}
\usepackage{float}

\usepackage{hyperref}
\usepackage[capitalise,nameinlink]{cleveref}
\usepackage[usenames,dvipsnames]{xcolor}
\hypersetup{
    colorlinks=true,       % false: boxed links; true: colored links
    linkcolor=Maroon,          % color of internal links (change box color with linkbordercolor)
    citecolor=OliveGreen,        % color of links to bibliography
    filecolor=magenta,      % color of file links
    urlcolor=Blue           % color of external links
}
\crefname{ansatz}{Ansatz.}{Ansatzes.}

\def\ket#1{\left|#1\right\rangle}
\def\bra#1{\langle#1|}

\begin{document}

\title{Coupled Cluster Green's function formulations based on the effective  Hamiltonians}

\author{Nicholas P. Bauman}
\email{nicholas.bauman@pnnl.gov}
 \affiliation{
        William R. Wiley Environmental Molecular Sciences Laboratory, Battelle, 
        Pacific Northwest National Laboratory, K8-91, P.O. Box 999, Richland WA 99352, USA}
\author{Bo Peng}
\email{peng398@pnnl.gov}
 \affiliation{
        William R. Wiley Environmental Molecular Sciences Laboratory, Battelle, 
        Pacific Northwest National Laboratory, K8-91, P.O. Box 999, Richland WA 99352, USA}
\author{Karol Kowalski} 
\email{karol.kowalski@pnnl.gov}
 \affiliation{
        William R. Wiley Environmental Molecular Sciences Laboratory, Battelle, 
        Pacific Northwest National Laboratory, K8-91, P.O. Box 999, Richland WA 99352, USA}
%\email{nicholas.bauman@pnnl.gov}
%\email{karol.kowalski@pnnl.gov}
%\email{peng398@pnnl.gov}

%%%\maketitle
%
% QUESTIONS TO ANSWER
%
% bare H in active space: tilde-T_int - how was %calculated?I presume that in the same way as for 
%the DUCC Hamiltonains
%
% Geomtries of N2 and H2O
%
\begin{abstract} 
We demonstrate that the  effective Hamiltonians obtained with the downfolding procedure based on double unitary coupled cluster (DUCC) ansatz can be  used in the context of Green's function coupled cluster (GFCC) formalism to calculate spectral functions of molecular systems. 
This combined approach (DUCC-GFCC) provides a significant reduction of numerical effort and 
good agreement with the corresponding all-orbital GFCC methods 
in energy windows that are consistent with the choice of active space.
These features are demonstrated on the example of two benchmark systems: H$_2$O and N$_2$, where DUCC-GFCC calculations were performed for active spaces of various sizes. 
\end{abstract}

\maketitle

\section{Introduction}
Although the Green's function formalism has been introduced as a tool to analyze properties of molecular systems more than five decades ago,\cite{hedin65_a796,cederbaum75_290,schirmer82_2395,cederbaum77_4124,book81,schirmer83_1237,cederbaum84_57,mukherjee89_5578,cederbaum92_2790,cederbaum96_7122,schirmer99_9982,rehr00_621,ortiz13_123,ortiz97,ortiz98_1008,schilfgaarde06_226402,louie06_216405,louie11_186404} it remains an active area of development. 
Recently, a considerable effort has been expended to provide 
accurate representations of the Green's function or self-energies for many-electron systems  in situations where a detailed characterization of correlation effects in molecular systems is necessary to achieve the required level of accuracy. 
\cite{chan16_235139,zgid14_241101,zgid12_165128,shee2019coupled,zhu2019coupled}
This includes a significant progress achieved in the development of (1) various diagrammatic techniques and linked cluster theorems for perturbative expansions,\cite{hirata2015general,hirata17_044108} (2)  stochastic approaches to evaluate the self energies,\cite{johnson2018monte} (3) cumulant representation,\cite{kas14_085112,kas2015real} and (4) explicitly correlated formulations \cite{johnson2018monte,pavovsevic2017communication} to calculate self-energies or Green's functions of molecular systems. 
Another important impulse to developed high-accuracy representation of Green's function/self-energy is associated with the development of more accurate embedding techniques related to dynamical mean-field theories (DMFT) \cite{zgid11_094115,zgid12_165128}
and self-energy embedding theory (SEET),\cite{zgid15_241102,zgid17_2200} where the  utilization of high-accuracy self-energies has been recently tested in studies of molecular systems. The accurate Green's function algorithms have been developed for conventional computer architectures and more recently for quantum computing. 
\cite{bauer2016hybrid,kosugi2019construction}
%embedding methods

A very promising way of introducing correlation effects to Green's function theory stems from the utilization of accurate coupled cluster (CC) expansions for the ground-state wave function.\cite{coester58_421,coester60_477,cizek66_4256,paldus72_50,purvis82_1910,paldus1999critical,crawford2000introduction,bartlett07_291,shavitt2009many} 
The coupled cluster Green's function formulation originally introduced by Nooijen and Snijders in a series of papers published in the early nineties 
\cite{nooijen92_55, nooijen93_15, nooijen95_1681}
(see also Ref. \cite{meissner93_67})
has been recently re-adopted by several groups in studies of atomic/molecular,\cite{kowalski14_094102,kowalski18_4335,matsushita18_224103,matsushita18_034106,lange2018relation,berkelbach2018communication}  and condensed phase and periodic systems,\cite{chan16_235139,matsushita18_204109} where the efficiency of models built upon 
inclusion hierarchy of collective many-body effect has been examined. 

Recent progress in CC theory led to the emergence of techniques for downfolding or dimensionality reduction of the electronic Hamiltonians. These techniques draw heavily on the utilization  of a tailored  double unitary coupled cluster (DUCC) expansion to integrate out high-energy wave-function components from low-energy ones in the effective (or downfolded) Hamiltonians. 
In this paper, we examine a variant of the GFCC formulations that utilize this class of effective Hamiltonians. In particular, our focus is on the active-space GFCC formulations that utilize standard GFCC models that build upon single, double (GFCCSD)\cite{nooijen92_55, nooijen93_15, nooijen95_1681,kowalski18_4335}  and single, double, and internal triple excitations (the so-called GFCC-i(2,3) approximation of Ref. \cite{peng2018green}). 
Using water molecule and N$_2$ systems as benchmarks, we demonstrate the efficiency of using downfolded Hamiltonians in recovering all features of spectral functions obtained in full GFCCSD and GFCC-i(2,3) calculations in energy windows that correspond to the active space used.

\section{Theory}
The main idea behind the DUCC-GFCC formalism is to use DUCC downfolded/effective Hamiltonians in the   GFCC calculations limited to the active space employed in the DUCC ansatz. Below, we describe the DUCC formalism, GFCC methodology, and combined DUCC-GFCC framework. 
\subsection{DUCC formalism}
In Ref.~\cite{bauman2019downfolding}, we introduced the unitary extension of the sub-system embedding sub-algebra CC approach (SES-CC)~\cite{safkk} which utilizes the double unitary CC expansion
\begin{equation}
	\ket{\Psi}=e^{\sigma_{\rm ext}} e^{\sigma_{\rm int}}\ket{\Phi}.
\label[ansatz]{ducc1}
\end{equation}
The character of the expansion (\ref{ducc1})  is similar to the  expansion  discussed in the 
single-reference formulation of the active-space coupled cluster formalism,\cite{active3, activerev}  
(see Refs. \cite{ active1, active2})  which also utilizes the decomposition of the cluster operator into internal and external parts. Additionally, using the {\it Campbell-Baker-Hausdorff} formula, 
expansion (\ref{ducc1}) can be viewed as yet another unitary CC ansatz with a specific form of unitary cluster amplitudes  (see Eq. (47) in  Ref. \cite{bauman2019downfolding})

In analogy to Ref.~\cite{bauman2019downfolding},  $\sigma_{\rm int}$ and $\sigma_{\rm ext}$ are the anti-Hermitian operators 
($\sigma_{\rm int}^{\dagger}=-\sigma_{\rm int}$ and $\sigma_{\rm ext}^{\dagger}=-\sigma_{\rm ext}$)
defined by excitations/de-excitations within and outside of active space, respectively. To be more 
precise, the amplitudes defining the $\sigma_{\rm ext}$ operator must carry at least one inactive spin-orbital index. 
Using~\cref{ducc1} in Schr\"odinger's equation, one obtains equations for cluster amplitudes and the 
corresponding energy, i.e., 
\begin{align}
        Qe^{-\sigma_{\rm int}}e^{-\sigma_{\rm ext}} H e^{\sigma_{\rm ext}}e^{\sigma_{\rm int}} \ket{\Phi} &= 0,
\label{uccd2eq} \\
        \bra{\Phi}e^{-\sigma_{\rm int}}e^{-\sigma_{\rm ext}} H e^{\sigma_{\rm ext}}e^{\sigma_{\rm int}} \ket{\Phi} &= E,
\label{uccd2ene}
\end{align}
where $Q$ is a projection operator on the space spanned by determinants that are orthogonal to the reference function $\ket{\Phi}$.
In these and subsequent equations, we consider the case of the exact limit ($\sigma_{\rm int}$ and 
$\sigma_{\rm ext}$ include all possible excitations). 
In Ref.~\cite{bauman2019downfolding}, we showed that when $\sigma_{\rm int}$ contains all possible excitations/de-excitations within the 
active space, the energy of the system~\cref{uccd2ene} can be obtained by diagonalizing
the DUCC effective Hamiltonian
\begin{equation}
        \overline{H}_{\rm ext}^{\rm eff(DUCC)} e^{\sigma_{\rm int}} \ket{\Phi} = E e^{\sigma_{\rm int}}\ket{\Phi},
\label{duccstep2}
\end{equation}
where
\begin{equation}
        \overline{H}_{\rm ext}^{\rm eff(DUCC)} = (P+Q_{\rm int}) \overline{H}_{\rm ext}^{\rm DUCC} (P+Q_{\rm int})
\label{equivducc}
\end{equation}
and 
\begin{equation}
        \overline{H}_{\rm ext}^{\rm DUCC} =e^{-\sigma_{\rm ext}}H e^{\sigma_{\rm ext}}.
\label{duccexth}
\end{equation}
In the above eigenvalue  problem, the   $e^{\sigma_{\rm int}}\ket{\Phi}$   expansion defines the corresponding eigenvector and  $P$ and $Q_{\rm int}$ are  projection operators onto the reference function $|\Phi\rangle$ and excited determinants in the active space that orthogonal to $\ket{\Phi}$, respectively.  

To prove this property, it is sufficient to introduce the resolution of identity $e^{\sigma_{\rm 
int}}e^{-\sigma_{\rm int}}$ to the left of the $\overline{H}_{\rm ext}^{\rm DUCC} $ operator 
in 
\begin{equation}
	(P+Q_{\rm int}) \overline{H}_{\rm ext}^{\rm DUCC} e^{\sigma_{\rm int}} |\Phi\rangle = E
    (P+Q_{\rm int}) e^{\sigma_{\rm int}}|\Phi\rangle\;,
\label{duccstep1}
\end{equation}
where we employed the fact that 
\begin{equation}
(P+Q_{\rm int}) e^{\sigma_{\rm int}}|\Phi\rangle =
e^{\sigma_{\rm int}}|\Phi\rangle \;,
\label{pqipro}
\end{equation}
and to notice that $e^{-\sigma_{\rm int}}\overline{H}_{\rm ext}^{\rm 
DUCC} e^{\sigma_{\rm int}}=e^{-\sigma_{\rm int}}e^{-\sigma_{\rm ext}} H e^{\sigma_{\rm 
ext}}e^{\sigma_{\rm int}}$. 
%Next, in analogy to Eqs. (\ref{analysis1}) and 
%(\ref{llineq}), Eq. (\ref{equivducc}) can be represented as 
Next, using matrix representation of the $\sigma_{\rm int}$ operator in the CAS space, denoted 
as $\bm{\sigma}_{\rm int}$, this equation can be re-written as 
\begin{equation}
	[e^{\bm{\sigma}_{\rm int}}] [\bm{y}] = 0 \;,
\label{llineq2}
\end{equation}
where the first component of the $[\bm{y}]$ vector is equivalent to 
$\langle\Phi|e^{-\sigma_{\rm int}}e^{-\sigma_{\rm ext}} H 
e^{\sigma_{\rm ext}}e^{\sigma_{\rm int}} |\Phi\rangle-E$
while the remaining components correspond to projections of 
$e^{-\sigma_{\rm int}}e^{-\sigma_{\rm ext}} H 
e^{\sigma_{\rm ext}}e^{\sigma_{\rm int}} |\Phi\rangle$
onto excited configurations belonging to $Q_{\rm int}$. 
The $[e^{\bm{\sigma}_{\rm int}}]$ matrix is also  non-singular, which is a consequence of the formula 
\begin{equation}
	{\rm det}(e^{\bm{\sigma}_{\rm int}})=e^{{\rm Tr}(\bm{\sigma}_{\rm int})} =1
\label{det1}
\end{equation}
and the anti-Hermitian character of the $\bm{\sigma}_{\rm int}$
matrix, i.e., ${\rm Tr}(\bm{\sigma}_{\rm int})=0$
(where real character of $\sigma_{\rm int}$ cluster amplitudes is assumed).
Given the non-singular character 
of the $[e^{\bm{\sigma}_{\rm int}}]$ matrix 
(see also  Ref.~\cite{bauman2019downfolding}), this proves the equivalence of these two representations.

\subsection{GFCC methodology}
Our GFCC implementations follow the basic tenets of the original GFCC formalism introduced by Nooijen {\em et al.}\cite{nooijen92_55, nooijen93_15, nooijen95_1681} 
and its features discussed  in Refs. \cite{kowalski14_094102,kowalski16_144101,kowalski16_062512,kowalski18_561,kowalski18_4335,peng2018green,peng2019approximate}.
Using  CC bi-variational approach, the corresponding frequency-dependent Green's function for an $N$-particle system can be expressed as
\begin{eqnarray}
&&G_{pq}(\omega) =  \notag \\
&&\langle\Phi|(1+\Lambda)e^{-T} a_q^{\dagger} (\omega+(H-E_0)- \text{i} \eta)^{-1} a_p e^T |\Phi\rangle + \notag \\
&& \langle\Phi|(1+\Lambda)e^{-T} a_p (\omega-(H-E_0)+ \text{i} \eta)^{-1} a_q^{\dagger} e^T |\Phi\rangle \;. 
\label{gfxn0}
\end{eqnarray}
where the $a_p$ ($a_p^{\dagger}$) operator is the annihilation (creation) operator for an electron in  the $p$-th spin orbital. The $\omega$ parameter denotes the frequency, and the imaginary part $\eta$ is often called a broadening factor. The cluster operator $T$ and de-excitation operator $\Lambda$ define  correlated ket ($|\Psi\rangle$) and bra 
($\langle\Psi|$)
ground-state wave functions for 
for $N$-electron system 
\begin{eqnarray}
|\Psi\rangle&=&e^T |\Phi\rangle \;,
\label{ccfun} \\
\langle\Psi| &=& \langle\Phi|(1+\Lambda)e^{-T} \;.
\label{ccfunl}
\end{eqnarray}
The ground-state energy $E_0$, and the amplitudes defining $T$ and $\Lambda$ operators are obtained from the following sequence of CC equations
\begin{eqnarray}
Q e^{-T}He^T|\Phi\rangle &=& 0 ~, \label{eq:cceq} \\
\langle\Phi|e^{-T}He^T|\Phi\rangle &=& E_0 ~, \label{eeq} \\
\langle\Phi|(1+\Lambda) e^{-T}He^T Q &=& E_0 \langle \Phi|(1+\Lambda)Q ~, \label{eql} 
\end{eqnarray}
where the $T$ and $\Lambda$ operators are defined as
\begin{eqnarray}
T &=& \sum_{n=1}^{N}
	\frac{1}{(n!)^2}\sum_{\substack{i_1,\ldots,i_n;\\ a_1,\ldots, a_n}} 
	t^{i_1\ldots i_n}_{a_1\ldots a_n} 
	a^{\dagger}_{a_1}\ldots a^{\dagger}_{a_n} 
	a_{i_n}\ldots a_{i_1} \label{tn} \;,\\
\Lambda &=& \sum_{n=1}^{N}
	\frac{1}{(n!)^2}\sum_{\substack{i_1,\ldots,i_n;\\ a_1,\ldots, a_n}} 
	\lambda_{i_1 \ldots i_n}^{a_1\ldots a_n} 
	a^{\dagger}_{i_1} \ldots a^{\dagger}_{i_n} 
	a_{a_n}\ldots a_{a_1} \;, 
\end{eqnarray}
with $t^{i_1\ldots i_n}_{a_1\ldots a_n}$ and $\lambda_{i_1 \ldots i_n}^{a_1\ldots a_n}$ being the antisymmetric amplitudes, and the indices $i,j,k,\ldots$ ($i_1,i_2, \ldots$)  and $a,b,c,\ldots$ ($a_1, a_2,\ldots$) corresponding to occupied and unoccupied spin orbitals in the reference function $|\Phi\rangle$, respectively. The projection operator  $Q$ is defined as,
\begin{equation}
Q=\sum_{n=1}^N \frac{1}{(n!)^2} \sum_{\substack{i_1,\ldots,i_n;\\ a_1,\ldots, a_n}} |\Phi_{i_1\ldots i_n}^{a_1\ldots a_n}\rangle \langle\Phi_{i_1\ldots i_n}^{a_1\ldots a_n}|\;,
\label{qproj}
\end{equation}
and represents the projection onto the subspace spanned by  excited configurations $|\Phi_{i_1\ldots i_n}^{a_1\ldots a_n}\rangle$ defined as  $a_{a_1}^{\dagger} \ldots a_{a_n}^{\dagger}  a_{i_n}\ldots a_{i_1} |\Phi\rangle$.

Using the resolution of identity $\mathds{1}=e^{-T}e^T$, the algebraic expression for matrix elements of the Green's function can be re-written as
%and the similarity transformation to the Hamiltonian $H$, the annihilation operator $a_p$, and the creation operator $a_q^\dagger$, Eq. (\ref{gfxn0}) can be further expressed as
%
\begin{eqnarray}
G_{pq}(\omega) = 
&&\langle\Phi|(1+\Lambda) \overline{a_q^{\dagger}} (\omega+\overline{H}_N- \text{i} \eta)^{-1} 
	\overline{a_p} |\Phi\rangle + \notag \\
&& \langle\Phi|(1+\Lambda) \overline{a_p} (\omega-\overline{H}_N+ \text{i} \eta)^{-1} 
	\overline{a_q^{\dagger}} |\Phi\rangle \;,
\label{gfxn1}
\end{eqnarray}
where the similarity transformed operator $\overline{H}_N$ (in normal  product form representation), $\overline{a_p}$, and $\overline{a_q^\dagger}$ are defined as
\begin{eqnarray}
\overline{H}_N &=& e^{-T} H ~e^{T} - E_0, \\
\overline{a_p} &=& e^{-T} a_p ~e^{T}, \label{ap} \\
\overline{a_q^\dagger} &=& e^{-T} a_q^\dagger ~e^{T}. \label{aq} 
\end{eqnarray}
Note that by using {\it Campbell-Baker-Hausdorff} formula
\begin{equation}
e^{-B}Ae^B=A+[A,B]+\frac{1}{2} [[A,B],B]+ \ldots
\label{ch}
\end{equation}
one can derive the explicit forms of Eqs. (\ref{ap}) and (\ref{aq}) where
\begin{eqnarray}
\overline{a_p} &=& a_p+[a_p,T] \label{shortaim} \\
\overline{a_q^{\dagger} }&=& a_q^{\dagger}+[a_q^{\dagger},T] \label{shortaid}.
\end{eqnarray}
Now we  define  frequency dependent ionization-potential equation-of-motion CC (IP-EOMCC)  type operators $X_p(\omega)$  and 
frequency dependent electron-affinity EOMCC (EA-EOMCC)  type operators $Y_q(\omega)$
%on an {\it (N-1)}-particle Hilbert space
%
\begin{eqnarray}
X_p(\omega) &=& \sum_{i} x^i(p, \omega)  a_i  + \sum_{i<j,a} x^{ij}_a(p, \omega) a_a^{\dagger} a_j a_i +\ldots ~, 
\label{xp}  \\
Y_q(\omega) &=& \sum_{i} y_a(q, \omega) a_a^\dagger  + \sum_{i,a<b} y^{i}_{ab}(q, \omega) a_a^{\dagger} a_b^\dagger a_i +\ldots ~,  \ 
 \label{yp}
\end{eqnarray}
defined as
\begin{eqnarray}
(\omega+\overline{H}_N - \text{i} \eta )X_p(\omega)|\Phi\rangle = 
	\overline{a}_p |\Phi\rangle \;, \label{eq:xp} \\
(\omega-\overline{H}_N + \text{i} \eta )Y_q(\omega)|\Phi\rangle = 
	\overline{a_q^\dagger} |\Phi\rangle \;.\label{eq:yq}
\end{eqnarray}
Note that the amplitudes $x^i(p,\omega)$, $x^{ij}_a(p, \omega)$, $\cdots$, and $y_a(q, \omega)$, $y^{i}_{ab}(q, \omega)$, $\cdots$ are defined on  the entire  complex plane, and are functions of a spin-orbital index and frequency.
Consequently, we can express Eq. (\ref{gfxn1}) in a compact form
\begin{eqnarray}
G_{pq}(\omega) = 
&&\langle\Phi|(1+\Lambda) \overline{a_q^{\dagger}} X_p(\omega) |\Phi\rangle + \notag \\
&& \langle\Phi|(1+\Lambda) \overline{a_p} Y_q(\omega) |\Phi\rangle \;.
\label{gfxn2}
\end{eqnarray}
\subsection{GFCC methodology utilizing DUCC effective Hamiltonians}

For simplicity, let designate the second quantized representation of the $H^{{\rm eff}(DUCC)}_{\rm ext}$ operator by $\Gamma$. We will also consider the case when 
the set of active orbitals consists of all occupied orbitals and a small subset of active virtual orbitals (containing $n_{v}^{\rm act}$ active virtual orbitals), where, in general, $n_{v}^{\rm act}\ll n_v$, where $n_v$ designates the total number of virtual orbitals.

The main idea behind the combined GFCC and DUCC formalism (DUCC-GFCC) is to replace $T$, $\Lambda$, and $H$ operators in Eq. (\ref{gfxn0}) by  cluster, de-excitation and $\Gamma$ operators acting in the active space only. To 
avoid possible conflict with the notational convention used in Ref.\cite{...} to define the  $\sigma_{\rm int}$ operator, we will denote the active-space counterparts of the 
$T$ and $\Lambda$ operators by $\widetilde{T}_{\rm int}$ and $\widetilde{\Lambda}_{\rm int}$, respectively. 
The equations (\ref{eq:cceq})-(\ref{eql})
are replaced by their "active" counterparts
%\begin{widetext}
\begin{eqnarray}
Q_{\rm int} e^{-\widetilde{T}_{\rm int}}\Gamma 
e^{\widetilde{T}_{\rm int}}|\Phi\rangle &=& 0 ~, \label{eq:cceqa} \\
\langle\Phi|e^{-\widetilde{T}_{\rm int}}\Gamma 
e^{\widetilde{T}_{\rm int}}|\Phi\rangle &=& E_0^{\rm int} ~, \label{eeqa} \\
\langle\Phi|(1+\widetilde{\Lambda}_{\rm int}) e^{-\widetilde{T}_{\rm int}}\Gamma e^{\widetilde{T}_{\rm int}} Q_{\rm int} &=& E_0^{\rm int} \langle \Phi|(1+\widetilde{\Lambda}_{\rm int})Q_{\rm int} ~, \label{eqla} 
\end{eqnarray}
%\end{widetext}
Now, the coupled cluster Green's function employing  the DUCC Hamiltonian $\Gamma$ can be expressed as for active orbitals as follows
%\begin{widetext}
\begin{eqnarray}
G_{PQ}^{\rm DUCC}(\omega) &=& \langle\Phi|(1+\widetilde{\Lambda}_{\rm int})e^{-\widetilde{T}_{\rm int}} a_Q^{\dagger} (\omega+(\Gamma-E_0^{\rm int})- \text{i} \eta)^{-1} a_P e^{\widetilde{T}_{\rm int}} |\Phi\rangle + \notag \\
&& \langle\Phi|(1+\widetilde{\Lambda}_{\rm int})e^{-\widetilde{T}_{\rm int}} a_P (\omega-(\Gamma-E_0^{\rm int})+ \text{i} \eta)^{-1} a_Q^{\dagger} e^{\widetilde{T}_{\rm int}} |\Phi\rangle \;, 
\label{gfxn0ac}
\end{eqnarray}
%\end{widetext}
where indices $P,Q,\ldots$ designate active spin orbitals. 
Again, applying the resolution of identity $e^{-\widetilde{T}_{\rm int}}e^{\widetilde{T}_{\rm int}}$ in 
the above equation, one gets the follwing expressions for DUCC Green's function matrix elements
%\begin{widetext}
\begin{eqnarray}
G_{PQ}^{\rm DUCC}(\omega) = 
&&\langle\Phi|(1+\widetilde{\Lambda}_{\rm int}) \overline{a_Q^{\dagger}}^{\rm int} (\omega+\overline{\Gamma}_N- \text{i} \eta)^{-1} 
	\overline{a_P}^{\rm int} |\Phi\rangle + \notag \\
&& \langle\Phi|(1+\widetilde{\Lambda}_{\rm int}) \overline{a_P}^{\rm int} (\omega-\overline{\Gamma}_N+ \text{i} \eta)^{-1} 
	\overline{a_Q^{\dagger}}^{\rm int} |\Phi\rangle \;,
\label{gfxn1ints}
\end{eqnarray}
%\end{widetext}
where we used the following definitions
\begin{eqnarray}
\overline{\Gamma} &=& e^{-\widetilde{T}_{\rm int}} \Gamma ~e^{\widetilde{T}_{\rm int}} \;, \label{gbarac} \\
\overline{\Gamma}_N &=& \overline{\Gamma}-E_0^{\rm int}  \;,\label{gnomac} \\
\overline{a_P}^{\rm int} &=& e^{-\widetilde{T}_{\rm int}} a_P ~e^{\widetilde{T}_{\rm int}}, \label{apac} \\
\overline{a_Q^\dagger}^{\rm int} &=& e^{-\widetilde{T}_{\rm int}} a_Q^\dagger ~e^{\widetilde{T}_{\rm int}}. \label{aqac} 
\end{eqnarray}
In the  active-space driven DUCC-GFCC approach, the $X_p(\omega)$ and $Y_q(\omega)$
operators are replaced by $X_P^{\rm int}(\omega)$ and $Y_Q^{\rm int}(\omega)$, respectively, which are given by the following expressions:
\begin{widetext}
\begin{eqnarray}
X_P^{\rm int}(\omega) &=& \sum_{I} x^I(P, \omega)^{\rm int}  a_I  + \sum_{I<J,A} x^{IJ}_A(P, \omega)^{\rm int} a_A^{\dagger} a_J a_I +\ldots ~, \label{xpac}  \\
Y_Q^{\rm int}(\omega) &=& \sum_{A} y_A(Q, \omega)^{\rm int} a_A^\dagger  + \sum_{I,A<B} y^{I}_{AB}(Q, \omega)^{\rm int} a_A^{\dagger} a_B^\dagger a_I +\ldots ~,  
 \label{yqacp}
\end{eqnarray}
\end{widetext}
where indices $I,J,\ldots$ and $A,B,\ldots$ refer to active occupied and unoccupied spin orbitals indices, respectively (again, in the present discussion we assume that all occupied spin orbitals are treated as active). These operators  satisfy
\begin{eqnarray}
(\omega+\overline{\Gamma}_N - \text{i} \eta )X_P^{\rm int}(\omega)|\Phi\rangle = 
	\overline{a_P}^{\rm int} |\Phi\rangle \;, \label{eq:xpac} \\
(\omega-\overline{\Gamma}_N + \text{i} \eta )Y_Q^{\rm int}(\omega)|\Phi\rangle = 
	\overline{a_Q^\dagger}^{\rm int} |\Phi\rangle \;,\label{eq:yqac}
\end{eqnarray}
and the $G_{PQ}^{\rm DUCC}(\omega$) is given by the expression
%\begin{widetext}
\begin{eqnarray}
G_{PQ}^{\rm DUCC}(\omega) = 
&&\langle\Phi|(1+\Lambda_{\rm int}) \overline{a_Q^{\dagger}}^{\rm int}  X_P^{\rm int}(\omega) |\Phi\rangle + \notag \\
&& \langle\Phi|(1+\Lambda_{\rm int}) 
\overline{a_P}^{\rm int} Y_Q^{\rm int}(\omega) |\Phi\rangle \;.
\label{gfxn2}
\end{eqnarray}
%\end{widetext}

\section{Approximations and implementation details}
\label{sec_approx}
In the present study, we will consider the following approximations for the form of the $\Gamma$ operator and from of the $\widetilde{T}_{\rm int}$, $\widetilde{\Lambda}_{\rm int}$, $X_P^{\rm int}(\omega)$  and $Y_Q^{\rm int}(\omega)$ operators:
\begin{itemize}
    \item The $\Gamma$ operator only considers to by one- and two-body terms defined in Eq. (65) of Ref. \cite{bauman2019downfolding}. The $\Gamma$ operator is limited to the downfolding of inactive virtual orbitals.  
    \item $\widetilde{T}_{\rm int}$ and $\widetilde{\Lambda}_{\rm int}$ operators are  represented by single and double excitations within the active space (the $Q_{\rm int}$ operator of Eqs. (\ref{eq:cceqa}) and (\ref{eqla}) is replaced by the  projection operator onto singly and doubly excited configurations belonging to the active space $Q_{(1){\rm int}}$ and $Q_{(2){\rm int}}$, respectively). Moreover, we also assumed that $\widetilde{\Lambda}_{\rm int}=\widetilde{T}_{\rm int}^{\dagger}$ (the efficiency of this approximation was discussed in Refs. \cite{peng2018green}).
    \item $X_P^{\rm int}(\omega)$ and $Y_Q^{\rm int}(\omega)$ operators include single and double excitations with corresponding amplitudes defined by the active spin-orbital indices. Equations (\ref{eq:xpac})-(\ref{eq:yqac}) are projected onto the sub-spaces of $N-1$ and $N+1$ electron Hilbert spaces spanned by active-type single and double excitations
    ($Q^{N-1}_{(1){\rm int}}$, $Q^{N-1}_{(2){\rm int}}$ and $Q^{N+1}_{(1){\rm int}}$, $Q^{N+1}_{(2){\rm int}}$ respectively).
\end{itemize}
The implementations of  the DUCC-GFCCSD  formalism
described above
utilizes two existing computational components: (1) DUCC-Hamiltonian generator (\texttt{libDUCC} library \cite{bauman2019downfolding})
integrated with NWChem \cite{nwchem}) and (2) the parallel implementation of the GFCC formalism (\texttt{GFCCLib} library \cite{gfcclib}).
In the following discussion we will entirely focus on the 
ionization potential (IP) part of the DUCC CC Green's function, 
$G^{\rm  DUCC (IP)}_{PQ}(\omega)$
\begin{equation}
  G^{\rm DUCC(IP)}_{PQ}(\omega)=\langle\Phi|(1+\widetilde{\Lambda}_{\rm int}) \overline{a_Q^{\dagger}}^{\rm int}  X_P^{\rm int}(\omega) |\Phi\rangle
  \label{ipducc}
\end{equation}
and the corresponding spectral function $A(\omega)$
defined as
\begin{widetext}
\begin{equation}
A(\omega) = - \frac {1} {\pi} \text{Tr}_{\rm act} \left[ \Im\left({\bf G}^{\text{DUCC (IP)}}(\omega) \right) \right] 
\cong - \frac {1} {\pi} \sum_{I}^{occ} \Im\left(G_{II}^{\text{DUCC (IP)}}(\omega) \right)~.
\label{specf}
\end{equation}
\end{widetext}
%{\color{red} CHECK!summations}
where summation under trace ($\text{Tr}_{\rm act}$) includes summation over
active spin orbitals only.

Recently, we have introduced new class of GFCC approximations  where locations of poles are improved by extending the excitation level of inner auxiliary operators ($X_p(\omega)$, $Y_q(\omega)$).
\cite{peng2018green}
These approximations can be generally  categorized as GFCC-i(n, m) method, where the excitation level of the inner auxiliary operators (m) used to describe the ionization potentials and electron affinities effects in the $N - 1$ and $N + 1$ particle Hilbert spaces is higher than the excitation level (n) used to represent the ground-state coupled cluster wave function for the $N$-electron system. We also derived the so-called “n+1” rule  (or the GFCC-i(n,n+1) class of methods), which states that in order to maintain size-extensivity of the Green’s function matrix elements, the excitation level of inner auxiliary operators $X_p(\omega)$ and $Y_q(\omega)$ cannot be larger than n+1. We demonstrated that the GFCC-i(2,3) approximation can significantly improve the location of satellite peaks compared to the GFCCSD formalism.
\cite{peng2018green}
For this reason we decided to combine the GFCC-i(2,3) approach with the DUCC effective Hamiltonians (DUCC-GFCC-i(2,3) approach). 

The working equations for the DUCC-GFCC-i(2,3) method represent a simple modification of the GFCC-i(2,3) approach. For example, for the ionized part of the Green's functions considered here 
%we have chosen a form of Eq. (\ref{eq:xp23}), in which the projections on singles and doubles are same as in GFCCSDT (see Eqs. (\ref{x3_q1}) and (\ref{x3_q2}) where the full form of $X_{p,3}(\omega)$-dependent coupling terms including $Q_1^{(N-1)}V_N X_{p,3}(\omega)|\Phi\rangle$, $Q_2^{(N-1)}V_N X_{p,3}(\omega)|\Phi\rangle$, and $Q_2^{(N-1)}(V_N T_1 X_{p,3}(\omega))_C|\Phi\rangle$ is maintained).
%
\begin{widetext}
\begin{eqnarray}
Q_{\rm (1) int}^{(N-1)}(\omega+\overline{\Gamma}_N - \text{i} \eta ) 
(X_{P,1}^{\rm int}(\omega)+X_{P,2}^{\rm int}(\omega)+X_{P,3}^{\rm int}(\omega))|\Phi\rangle &=& Q_{\rm (1) int}^{(N-1)} \overline{a_P}^{\rm int}|\Phi\rangle~, \label{x3_q1} \\
Q_{\rm (2) int}^{(N-1)}(\omega+\overline{\Gamma}_N - \text{i} \eta ) 
(X_{P,1}^{\rm int}(\omega)+X_{P,2}^{\rm int}(\omega)+X_{P,3}^{\rm int}(\omega))|\Phi\rangle &=& Q_{\rm (2) int}^{(N-1)} \overline{a_P}^{\rm int}|\Phi\rangle~ \;, \label{x3_q2}  \\
Q_{\rm (3) int}^{(N-1)} \lbrace ((\omega+F_N(\Gamma)_D- \text{i} \eta) X_{P,3}^{\rm int}(\omega))_C  + (\overline{\Gamma}_{N,2} X_{P,2}^{\rm int}(\omega))_C + 
(\overline{\Gamma}_{N,2} \widetilde{T}_2^{\rm int} X_{P,1}^{\rm int}(\omega))_C \rbrace  |\Phi\rangle &=& Q_{\rm (3) int}^{(N-1)} \overline{a_P}^{\rm int}|\Phi\rangle  \;,  \notag  \label{x3_q3}
\end{eqnarray}
\end{widetext}
where the $F_N(\Gamma)_D$ is diagonal part of te one-particle  $\Gamma$-operator-dependent Fock operator in a normal product form,  $\overline{\Gamma}_{N,2}$ is  two-particle part of $\overline{\Gamma}_N$, and subscript "C" designates connected part of a given operator expression. 
The $X_{P,3}^{\rm int}(\omega)$ and $Y_Q^{\rm int}(\omega)$ are internal triply excited components of the $X_P^{\rm int}(\omega)$ and $Y_Q^{\rm int}(\omega)$ operators. 
One should mention that in analogy to the GFCC-i(2,3) case, the DUCC-GFCC-i(2,3) expression is given by DUCC-GFCCSD expression 
(Eq. (\ref{ipducc}), the only difference is in the fact that now the $X_{P,1}(\omega)$ and $X_{P,2}(\omega)$ are iterated in the presence of the $X_{P,3}(\omega)$ operator. 

%\newpage
\section{Results and Discussion}

\begin{figure}
\includegraphics[width=0.45\textwidth]{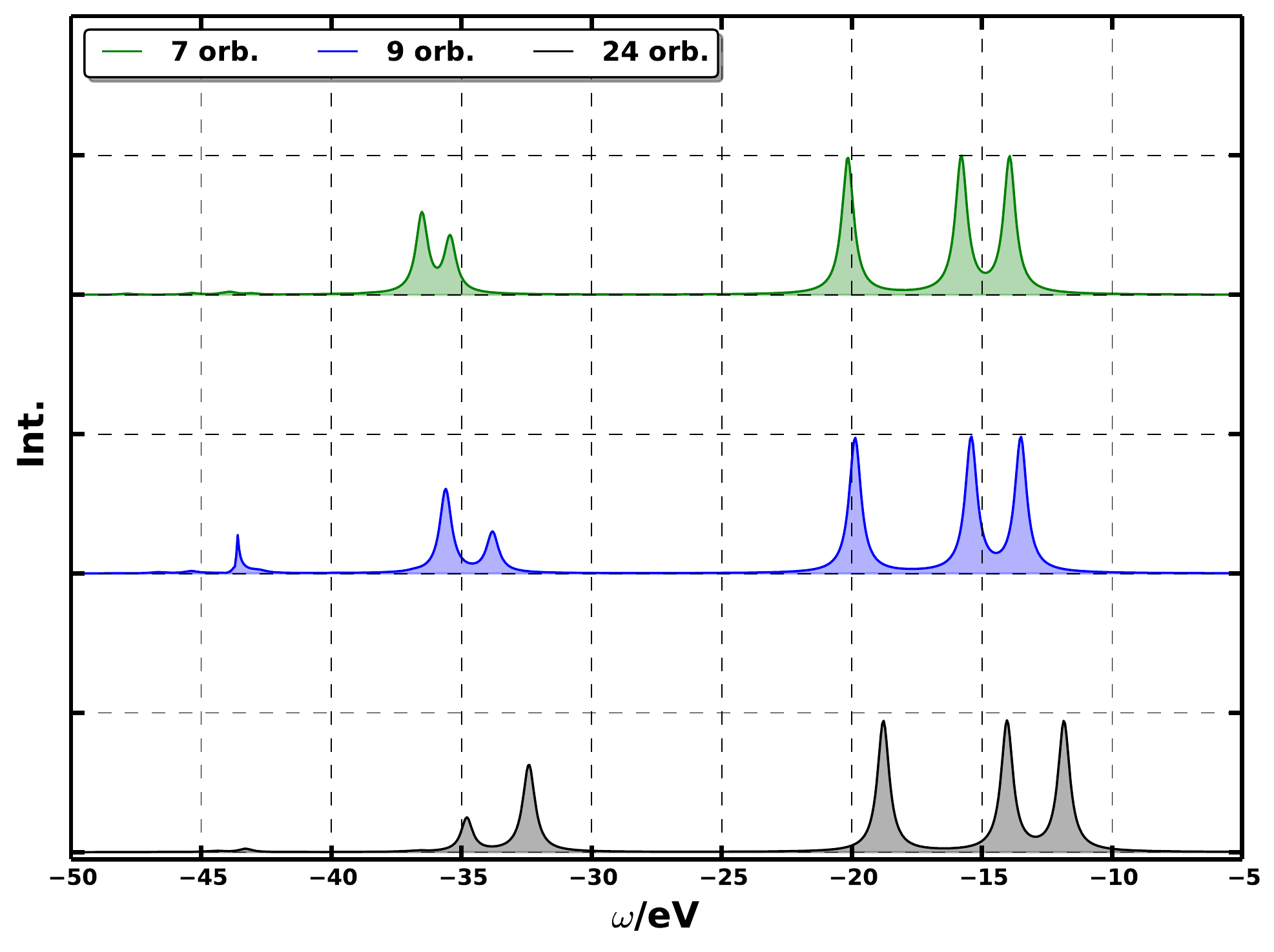}
\caption{Spectral functions of the water molecule in the valence energy regimes directly computed by the close-shell GFCCSD and DUCC-GFCCSD methods with cc-pVDZ basis set. The conventional GFCCSD results with 24 total number of molecular orbitals are shown as a black line at the bottom. The DUCC-GFCCSD results with 7 and 9 internal molecular orbitals (including all the five occupied molecular orbitals) are shown in green and blue lines, respectively.
\label{figh2o}}
\end{figure}

\begin{figure}
\includegraphics[width=0.45\textwidth]{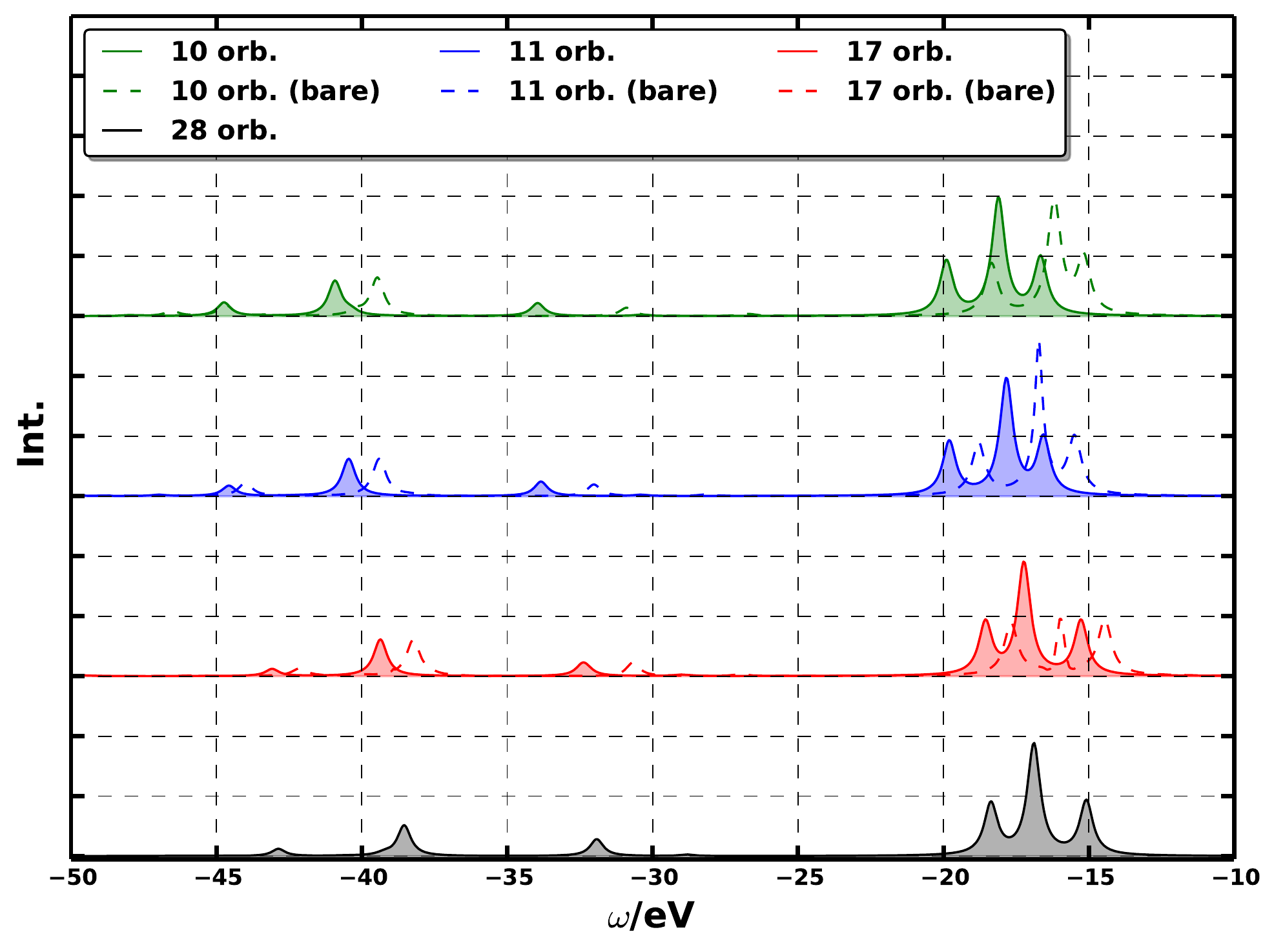}
\caption{Spectral functions of the nitrogen molecule in the valence energy regimes directly computed by the closed-shell GFCCSD and DUCC-GFCCSD methods with cc-pVDZ basis set. The conventional GFCCSD results with 28 total number of molecular orbitals are shown as a black line at the bottom. The DUCC-GFCCSD results with 10, 11, and 17 internal molecular orbitals (including all the seven occupied molecular orbitals) are shown in green, blue, and red lines, respectively, with their bare Hamiltonian counter parts being shown in the dashed line with same colors.
\label{fign2}}
\end{figure}

To illustrate the performance of the DUCC-GFCCSD formalism we performed  calculations for two benchmark systems: H$_2$O and N$_2$ molecules in cc-pVDZ basis sets.\cite{dunning89_1007} In both cases, we calculated spectral functions for DUCC Hamiltonians defined by active spaces of various sizes. For the water molecule, we tested two active spaces including 7 and 9 active orbitals (spanned by 5 occupied and 2  virtual orbitals and 5 occupied and 4 virtual orbitals, respectively). 
For N$_2$ molecules we used three active spaces composed of 10, 11, and 17 orbitals. For N$_2$ we have also performed GFCCSD calculations employing bare Hamiltonians in the same active spaces. 
The conventional  GFCCSD results employing all orbitals are shown in Figs. \ref{figh2o} and \ref{fign2} in black lines. 

It is intriguing to see that for both 
systems,  DUCC-GFCCSD spectral functions 
(see Figs. \ref{figh2o} and \ref{fign2}) approach monotonically conventional GFCCSD spectral functions as the size of active space is increased. Additionally, all features of the conventional GFCCSD formulation correlating all orbitals are reproduced by DUCC-GFCCSD approaches. This concerns not only the main peaks in the energy range [-20,-10] eV but also all three satellite peaks located below -30 eV.  

It is interesting to notice that the monotonic behavior of the DUCC-GFCCSD spectral functions is a consequence of the "correlated" character of the DUCC Hamiltonians, which also capture the dynamical correlation effects outside of active space. To illustrate this fact for the N$_2$ molecule we performed GFCCSD in the same active spaces using bare Hamiltonians, where the corresponding spectral functions are shown in Fig. \ref{fign2}
by dotted lines. In contrast to the DUCC-GFCCSD formalism, the spectral functions for GFCC approaches  utilizing bare Hamiltonian in active spaces  are no longer monotonically approaching full GFCCSD results. In some cases  the position of peaks disclose an irregular behavior. 
For example,  the first main peak of N$_2$ obtained with  11 active orbitals  (see Fig. \ref{fign2} blue line) overestimates the ionization energy obtained with the full GFCCSD approach whereas  for 17 active orbitals case 
(see Fig. \ref{fign2} red line) this quantity is underestimated. Similar behavior can be observed for the 
satellite peak of N$_2$ located between -30 and -35 eV.

\begin{figure}
\includegraphics[width=0.45\textwidth]{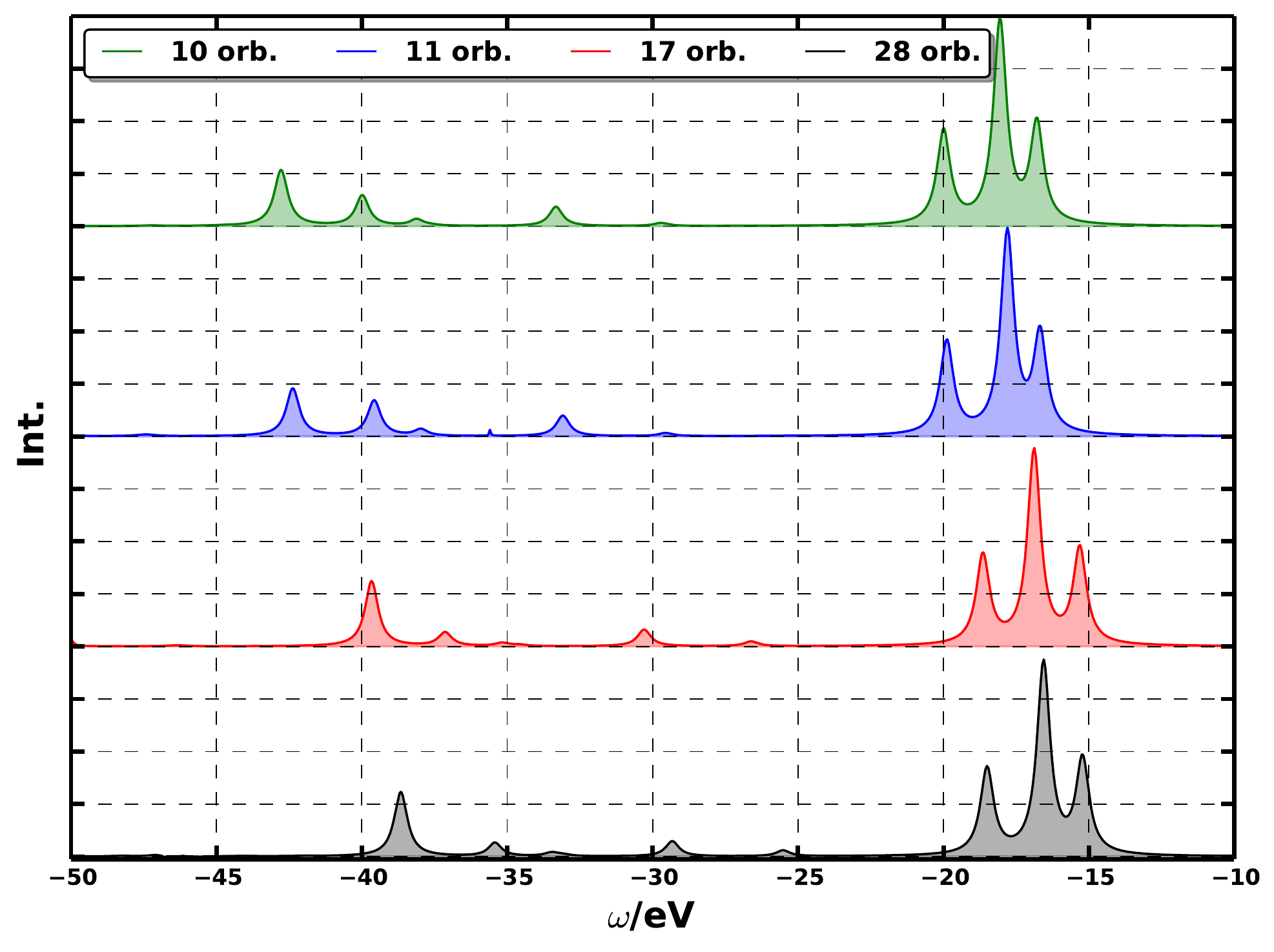}
\caption{Spectral functions of the nitrogen molecule in the valence energy regimes directly computed by the closed-shell GFCC-i(2,3) and DUCC-GFCC-i(2,3) methods with cc-pVDZ basis set. The conventional GFCC-i(2,3) results with 28 total number of molecular orbitals are shown as a black line at the bottom. The DUCC-GFCC-i(2,3) results with 10, 11, and 17 internal molecular orbitals (including all the seven occupied molecular orbitals) are shown in green, blue, and red lines, respectively
\label{fign3}}
\end{figure}

To asses to what extent DUCC-GFCC-i(2,3) formalism
can reproduce the effects  due to the
higher-order excitations (for example, triple excitation) in the calculated spectral functions, we performed the DUCC-GFCC-i(2,3) calculations for the N$_2$ system which was used in the original GFCC-i(2,3) studies. 
To make a comparison with the previous DUCC-GFCC calculations we used the same active spaces employed in the DUCC-GFCCSD case. 
Our results are shown in Fig. \ref{fign3}, where the exact GFCC-i(2,3) spectral function is marked in   gray. 
As one can see, all active spaces can reproduce all the basic features of the full GFCC-i(2,3) approach. It regards not only the main peaks but also all satellite peaks  (below -25 ev), of which two were not detected by the GFCCSD method. The results obtained with the 17 orbitals active space are in a good agreement with the full GFCC-i(2,3) results. One should emphasize that this agreement was possible to obtain with a very simple form of the effective Hamiltonians used in the simulations (see Eq. (65) of Ref. \cite{peng2018green}).

%% higher-order commutators in conclusions

\section{Conclusions}
We demonstrated that the utilization of the effective Hamiltonian stemming from the DUCC downfolding procedure can be used to reproduce the main features of the standard  GFCCSD spectral function. In a series of test calculations, we demonstrated that increasing active space size leads to monotonic improvements in the location of  peaks obtained with the  DUCC-GFCCSD approach with respect to the full GFCCSD results. We contribute this behavior to the presence of dynamical (out-of-active-space) correlation effects encapsulated in each of  DUCC effective Hamiltonians. In contrast to the DUCC-GFCCSD formalism, the utilization of active space bare Hamiltonians results in a less regular behavior of calculated peaks. 
The utilization of the DUCC effective Hamiltonians can also significantly reduce the cost of the GFCC calculations for the energy regime embraced by the corresponding active space.
We also demonstrated that the DUCC-GFCC-i(2,3) can encapsulate  necessary correlation effects needed for the description of satellite peaks that are not captured by lower rank GFCCSD and DUCC-GFCCSD formulations. 
Growing interest in the development of quantum computing algorithms for  correlated Green's function \cite{bauer2016hybrid,kosugi2019construction}  makes reduced-dimension DUCC-GFCC formulations a possible  target for early quantum computing applications.

\section{Acknowledgement}
This work was supported by the “Embedding Quantum Computing into Many-Body Frameworks for Strongly Correlated Molecular and Materials Systems” project, which is funded by the U.S. Department of Energy (DOE), Office of Science, Office of Basic Energy Sciences, the Division of Chemical Sciences, Geosciences, and Biosciences. A portion of this research was funded by the Quantum Algorithms, Software, and Architectures (QUASAR) Initiative at the Pacific Northwest National Laboratory (PNNL). It was conducted under the Laboratory Directed Research and Development Program at PNNL. Pacific Northwest National Laboratory is operated by Battelle for DOE under Contract DE-AC05-76RL01830.

%%\bibliography{gfcc}

%merlin.mbs apsrev4-1.bst 2010-07-25 4.21a (PWD, AO, DPC) hacked
%Control: key (0)
%Control: author (72) initials jnrlst
%Control: editor formatted (1) identically to author
%Control: production of article title (-1) disabled
%Control: page (0) single
%Control: year (1) truncated
%Control: production of eprint (0) enabled
%

\end{document}